\documentclass[prb,preprint,showpacs,preprintnumbers,amsmath,amssymb]{revtex4}

\usepackage{graphicx}
\usepackage{amsmath}
\usepackage{makeidx} 
\begin{document}

\title{
     	Two Modes of Magnetization Switching in a Simulated Iron Nanopillar in an Obliquely Oriented Field
      }

\author{
	S.H. Thompson,$^{1,2,3}$ G. Brown,$^{4,1}$ P.A. Rikvold,$^{1,2,3,5}$ M.A. Novotny$^{6,7}$
	}

\affiliation{
	$^1$Department of Physics,	\\
	Florida State University, Tallahassee, FL 32306-4350, USA\\
	$^2$School of Computational Science, \\
	Florida State University, Tallahassee, FL 32306-4120, USA\\
	$^3$Center for Materials Research and Technology, \\
	Florida State University, Tallahassee, FL 32306-4350, USA\\
	$^4$Center for Nanophase Materials Science,\\
	Oak Ridge, TN 37831-6164, USA \\
	$^5$National High Magnetic Field Laboratory, \\
	Tallahassee, FL 32310-3706, USA\\
	$^6$Department of Physics and Astronomy, \\ 
	Mississippi State University, Mississippi State, MS 39762, USA\\
	$^7$HPC$\,^2$ Center for Computational Sciences,\\
	Mississippi State University, Mississippi State, MS 39762, USA\\
}
\date{\today}

\begin{abstract}

Finite-temperature micromagnetics simulations are employed to study the magnetization-switching dynamics driven by a field applied at an angle to the long axis of an iron nanopillar.  
A bi-modal distribution in the switching times is observed, and evidence for two competing modes of magnetization-switching dynamics is presented.  
For the conditions studied here, temperature $T = 20$~K and the reversal field 3160 Oe at an angle of 75$^\circ$ to the long axis, approximately 70\% of the switches involve unstable decay (no free-energy barrier) and 30\% involve metastable decay (a free-energy barrier is crossed).
The latter are indistinguishable from switches which are constrained to start at a metastable free-energy minimum.
Competition between unstable and metastable decay could greatly complicate applications involving magnetization switches near the coercive field.

\end{abstract}

\pacs{
	75.75.+a, 
	75.60.Jk, 
	75.40.Mg, 
	85.70.Ay  
	}
	
\maketitle

\section{Introduction}

Nanoscale magnetic devices play important roles in many applications, including sensor technology and magnetic recording.
Fabricating such devices requires innovative techniques that rely on knowledge of the sensor's structure and dynamics at the nanoscale.
The role of a micromagnetic simulation in the design process, consequently, is to represent such systems with the appropriate resolution, i.e., preserving the actual physical dynamics while maintaining a reasonable simulation effort.
This has historically been achieved through the use of semi-classical equations that govern the motion of individual spins on a lattice mapped from a physical magnetic system.
In our work, the Landau-Lifshitz-Gilbert (LLG) equation~\cite{Brown:IEEE} provides such dynamics for the spins and includes finite-temperature effects by incorporating a stochastic field that obeys a fluctuation-dissipation relation.

Here we model the magnetization switching of a system with strong uniaxial shape anisotropy: an iron nanopillar.
When subjected to a magnetic field obliquely aligned with respect to the pillar axis, the pillar exhibits a bimodal distribution of switching times.~\cite{Thompson:rmsinoof} 
This is a feature which may have important ramifications for the application of such nanopillars in real-world devices, which typically rely on a single, consistent decay mode.
The traditional picture of a single free-energy barrier which must be crossed, often used to describe the decay of a metastable state, appears to be insufficient for this system.
Therefore, we analyze the bimodal switching behavior of simulated nanopillars using information from temporal phase portraits and also apply absorbing Markov-chain techniques~\cite{Novotny:atoadmcmfswdss, Iosifescu:fmpata, Kolesik:apmfsadolss, Kolesik:pdfmdiim, Brown:pdamr, Kolesik:eldmcs} to transition matrices obtained from the simulations.

The rest of this paper is organized as follows.
We first present our computational model and our implementation of the LLG equation in Section~\ref{MaNM}.
Section~\ref{NR} is divided into three parts, which collectively discuss results obtained from the simulations.
Section~\ref{PP} explores the phase portraits of the energy during switching, while Section~\ref{TM} and Section~\ref{PD} provide information about the free energy of the system based on analysis of transition matrices and projective dynamics, respectively.
Our conclusions are presented in Section~\ref{S}.

\section{Model and Numerical Method}
\label{MaNM}

Our numerical model is motivated by nanopillars fabricated by von Moln\'{a}r and collaborators.~\cite{Kent:science262, Wirth:mbonsip}
Using scanning-tunneling-microscopy-assisted chemical vapor deposition, they constructed elongated iron nanoparticles, each approximately $10\times10\times150$~$\rm{nm}^3$.
Their results indicated that the field-driven magnetization switching in these particles at field strengths near the coercive field is initiated by localized nucleation, followed by thermal activation over a free-energy barrier.~\cite{Wirth:famponspa, Li:mriefn, Wirth:tamrinsip, Li:hmoasin}
This is expected for systems whose dimensions are large enough to support a non-uniform magnetization,~\cite{Wieser:dwmintvvw, Hertel:mrdnn} in contrast to the single coherent rotation mode assumed by smaller particles via the Stoner-Wohlfarth model.~\cite{Stoner:amomhiha}  

In order to numerically investigate these nanopillars, we use a coarse-grained, cuboid computational lattice in which each cell represents the net magnetization of the corresponding volume in the physical system.
To ensure that the magnetization density is uniform at length scales below the cell volume, the lattice spacing is chosen smaller than the exchange length of $2.6$~nm, obtained from the material properties of bulk iron.~\cite{Brown:lsotamrinp}
This criterion yields a regular lattice with the dimensions $6\times6\times90$ ($N=3240$ spins, $\Delta{x} = 1.6667$~nm), which has a single classical Heisenberg spin at each site.
The time evolution of this spin, $\vec{m}(\vec{r_i})$, is controlled by the LLG equation,~\cite{Brown:IEEE, Aharoni:itf, Brown:m}
\begin{equation}
\frac{{\it{d}}{\vec{m}(\vec{r_i})}}{\it{dt}} = \frac{\gamma_0}{1+\alpha}\left(\vec{m}(\vec{r_i})\times 
\left[\vec{H}(\vec{r_i})-\frac{\alpha}{m_\mathrm{s}}\vec{m}(\vec{r_i})\times \vec{H}(\vec{r_i})\right]\right),
\label{eq:M}
\end{equation}
which updates every site on the lattice at each computational time step.
At each site $i$, the local field $\vec{H}(\vec{r_i})$ determines the direction of change of the magnetization $\vec{m}(\vec{r_i})$ during the next integration step.
The parameters $\gamma_0= 1.76 \times 10^7$~Hz/Oe, $m_\mathrm{s} = 1700$~$\mathrm{emu/cm^3}$, and $\alpha = 0.1$ represent the electronic gyromagnetic ratio, the saturation magnetization of bulk iron, and a phenomenological damping parameter, respectively.
Values for these parameters are consistent with the material properties of bulk iron, and are discussed in previous work.~\cite{Brown:lsotamrinp, Thompson:rmsinoof, Thompson:pdirmon}
The total local field, $\vec{H}(\vec{r_i})$, is a sum of the individual fields, which include the dipole field $\vec{H^\mathrm{D}}$, the applied field $\vec{H^\mathrm{Z}}$, the exchange field $\vec{H^\mathrm{E}}$, and the stochastic thermal field $\vec{H^\mathrm{T}}$.

Only nearest-neighbor interactions are considered for the exchange field $\vec{H^\mathrm{E}}$, while the long-range dipolar interaction, $\vec{H^\mathrm{D}}$, couples all computational cells to each other.
Consequently, the dipolar term constitutes the largest computational task during an integration step of the simulation, scaling as $O(N^2)$ for a brute force calculation.
This term is reduced to a linear dependence on the number of cells, $N$, by using the Fast Multipole Method,~\cite{greengard:treopfips, Brown:lsotamrinp} resulting in simulation rates of approximately $0.005$ ns/cpu-hour on a $3$~GHz Intel Pentium 4 processor with 512~MB of memory using the C++ Psimag library.~\cite{url:psimag}

The temperature is set to $20$ K and is included in the simulation through the stochastic field $\vec{H^{\mathrm{T}}}$, whose components are Gaussian distributed with mean zero and variance determined by the fluctuation-dissipation relation, 
\begin{equation}
{\langle}{H_{\beta}^\mathrm{T}}({\vec{r_i}},t){H_{\gamma}^\mathrm{T}}(\vec{r_j},t'){\rangle} = 
	\frac{2{\alpha}{k_\mathrm{B}}T}{{\gamma_0}M_\mathrm{s}{V}}\delta_{ij}\delta_{\beta\gamma}\delta(t-t'),
\end{equation}
where $k_\mathrm{B}$ is Boltzmann's constant, $V$ is the volume of an individual computational cell, $T$ is the absolute temperature, $\delta_{ij}$ and $\delta_{\beta\gamma}$ are Kronecker deltas over the lattice sites $i, j$ and directions $\beta, \gamma$, respectively, and $\delta(t-t')$ is a Dirac delta function of the time difference, $t-t'$.
This equation implies that the magnitude of the thermal field scales linearly with the square root of the temperature.
The details of the integration of the stochastic field are discussed in Ref.~\onlinecite{Brown:lsotamrinp}.

\begin{figure}[tb]
\centering\includegraphics[scale=0.40]{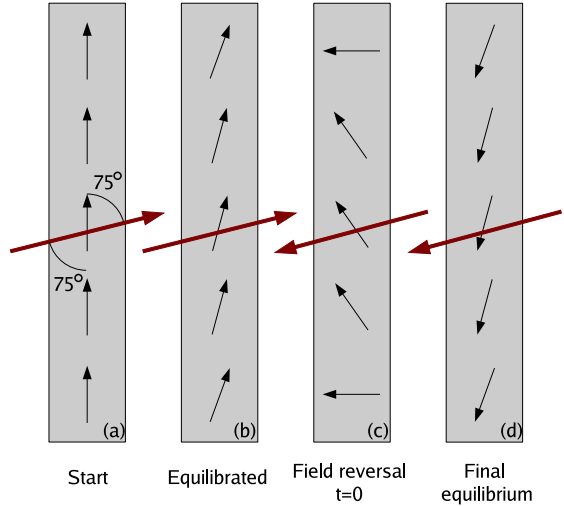}
\caption[]
{
(Color online.) Schematic of pillar magnetization (small arrows, black online) and the applied field (large arrows, red online) at different times during a trial.
(a) Initially, at $t=-0.25$~ns, the magnetization is aligned with the long axis of the pillar and the applied field has a value of $3160$~Oe, oriented at $75^\circ$ with respect to the long axis.
(b) Some time later, at $t=-0.125$~ns, the magnetization is relaxed and the applied field begins its sinusoidal reversal as described in the text.
(c) The field reversal is complete at $t=0$~ns.
(d) The final equilibrium of the simulation after magnetization switching has occurred.
}
\label{fig:InitTime}
\end{figure}

The simulation begins with the pillar in an applied field of magnitude $3160$~Oe and angle 75$^\circ$ with respect to the long axis of the pillar (Fig.\ref{fig:InitTime}(a)). 
Once equilibriated, the spins are relaxed in the direction of the initial applied field (Fig.\ref{fig:InitTime}(b)).
The value of the applied field is then changed sinusoidally to a direction anti-parallel to the initial field over a time period of $t=0.125$~ns, $\vec{H}(t) = \vec{H}(0)\cos(\pi{t}/0.125~\mathrm{ns})$, $t\in[-0.125~\mathrm{ns}, 0]$.
The $z$-component of $\vec{H}(0)$ is negative (Fig.\ref{fig:InitTime}(c))
This is done to ensure that excess energy is not artificially pumped into the system.
Upon completion of the field reversal, the simulation time is set to zero and the field is kept static at $\vec{H}(0)$ for the remainder of the trial.

\section{Numerical Results}
\label{NR}

\begin{figure}[tb]
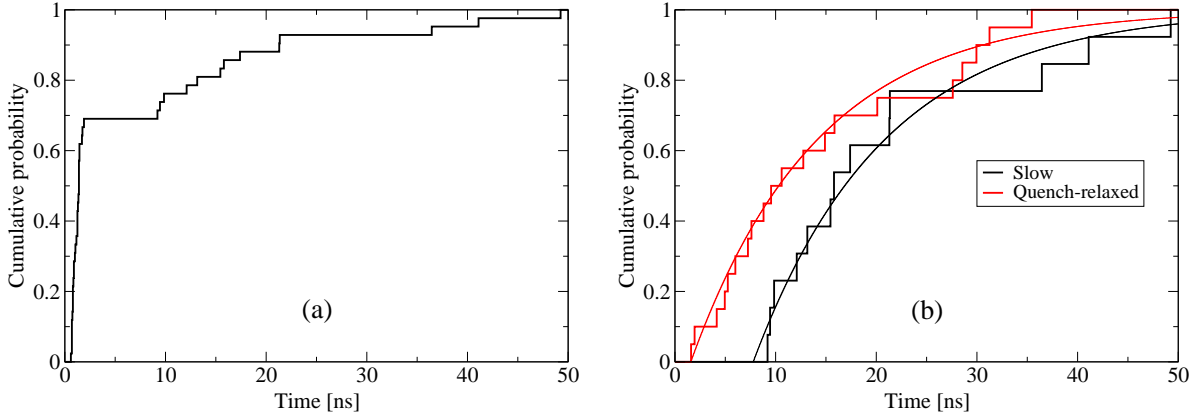

\begin{center}
$\begin{array}{c@{\hspace{0.5truecm}}c}
\includegraphics[scale=0.31]{2_DistributionFunctions_publication_75a} &
\includegraphics[scale=0.31]{3_DistributionFunctions_publication_75b}
\end{array}$
\end{center}
\caption[]
{
(Color online.) Cumulative switching-time distribution for all 42 full runs~(a) and a comparison of quenched-relaxed and slow runs (b). 
It is clear from (a) that at least two characteristic time scales are present in the switching statistics for this system.
From the measured lifetimes of the trials we obtain $\langle{t}\rangle = 1.2$ ns with a standard deviation $\sigma_{t} = 0.4$~ns for the fast mode ($29$ trials), and for the slow mode $\langle{t}\rangle = 20.9$~ns, $t_0 = 7.8$~ns, and $\tau = \sigma_{t} = 13.1$~ns ($13$ trials).
Also shown are the lifetimes for the QR trials with $\langle{t}\rangle = 14.2$~ns, $t_0 = 1.6$~ns, $\tau = 12.6$~ns and $\sigma_{t} = 10.8$~ns.
The fitted expressions, $D(t) = \eta(t-t_0)(1-\exp(-(t-t_{0})/{\tau}))$, where $\eta$ is the Heaviside step function, are also shown in~(b).
}
\label{fig:CD_1}
\end{figure}

Under the conditions described in Section~\ref{MaNM}, $42$ magnetization-switching simulations were performed.
The switching time $t_\mathrm{s}$ is defined as the first-passage time to $\sum_{i=1,N} m_{z}(\vec{r_i}) \leq 0$, with $m_{z}$ the component of the magnetization along the long axis of the pillar, and time measured from the completion of the initial field reversal.
From these $42$ trials, the cumulative distribution of the switching times, shown in Fig.~\ref{fig:CD_1}(a), indicates at least two characteristic time scales and the existence of more than a single switching path.
About $70$\% of the simulations (29 runs) switched almost immediately ($t_\mathrm{s}<2$~ns), while the remaining $30$\% (13 runs) exhibited switching times at least an order of magnitude larger.
Since direct comparison of the magnetization of individual runs did not reveal any differences in the switching mechanism, the trials were divided into two groups based only on the observed switching times: ``fast'' decay ($t_\mathrm{s}<2$~ns) and ``slow'' decay ($t_\mathrm{s}\geq2$~ns).
Below we show that the slow-mode statistics are the result of a process which must traverse a free-energy landscape characterized by a metastable well that the system must escape to reach the lowest available free-energy state.
This is accomplished by the collective effect of many random thermal fluctuations which can eventually cause the system to surmount the free-energy saddle point that separates the metastable well from the global free-energy minimum.

Thermally activated barrier-crossing rates from processes such as this are described by an Arrhenius-like expression and a switching-time probability density function (pdf) which assumes an exponential form.
To find the lifetime, $\tau$, of each mode, the mean switching time, $\langle{t}\rangle = (1/N)\sum_{i}t_i$, and the standard deviation,  $\sigma_t = \sqrt {(1/(N-1)) \sum_{i}(t_i-\langle{t}\rangle)^2}$, are determined from the empirical data.
The delayed exponential, $f(t) = \eta(t-t_0)(1/\tau)\exp(-(t-t_{0})/\tau)$, where $\eta$ is the Heaviside step function, and $t_{0} = \min\{$shortest switching time$, \langle{t}\rangle - \sigma_{t}\}$, is used as the pdf.
Once the waiting time $t_0$ is determined, the maximum likelihood estimate for the lifetime is found as $\tau = \langle{t}\rangle-t_{0}$.
The slow-mode lifetime from this estimate is $\tau = 13.1$~ns, with a corresponding $t_{0} = 7.8$~ns.
These results are shown in Fig.~\ref{fig:CD_1}(b) and Table~\ref{table1}, with $\tau$ for the slow mode close to other estimates that will be discussed below.
In comparison, the empirical mean for the fast mode is $\langle{t}\rangle = 1.2$~ns, with $\sigma_{t} = 0.4$~ns.
The corresponding pdf is not well approximated by a delayed exponential for the fast mode.

In addition to the above switches, $20$ separate simulations were also completed that are identical to the previous conditions except for the details of the initial magnetization and field-reversal.
The initial magnetization of these runs was determined by quenching the system to $0$ K while in the metastable well, and then re-thermalizing it to $20$ K.
From this time on ($t=0$), the simulations were carried out identically to the previous $42$ trials.
The reasons for this procedure are explained in Section~\ref{PP}.
Along with the slow mode, the cumulative distribution of lifetimes for the quenched-relaxed (QR) runs is shown in Fig.~\ref{fig:CD_1}(b), with $\tau = 12.6$ ns, $\sigma_t = 10.8$ ns, and $t_0 = 1.6$ ns.

\subsection{Phase Portraits}
\label{PP}

\begin{figure}[h]
\begin{center}
$\begin{array}{c@{\hspace{0.0truecm}}c@{\hspace{-0.0truecm}}c}
\includegraphics[scale=0.17]{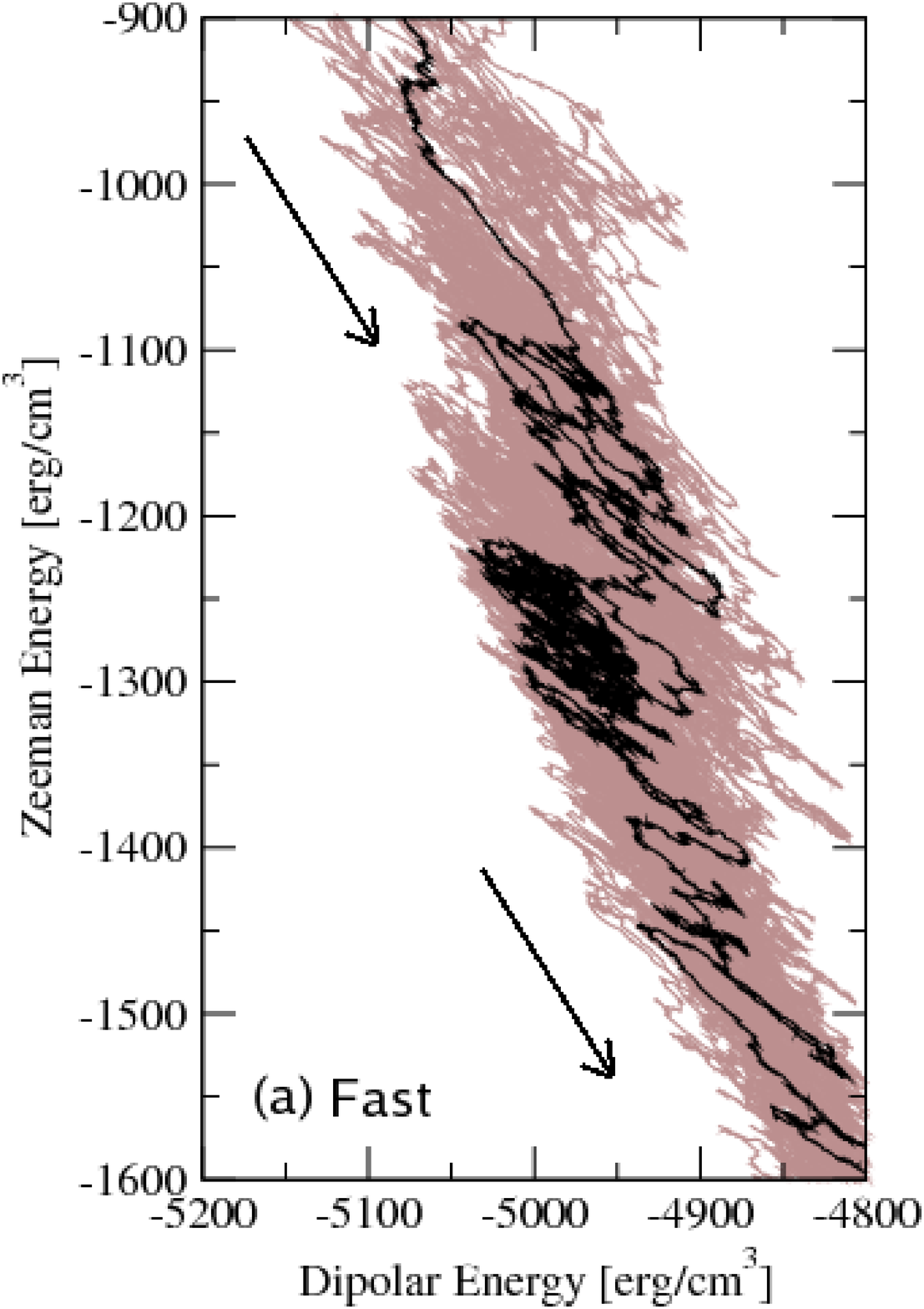} &
\includegraphics[scale=0.17]{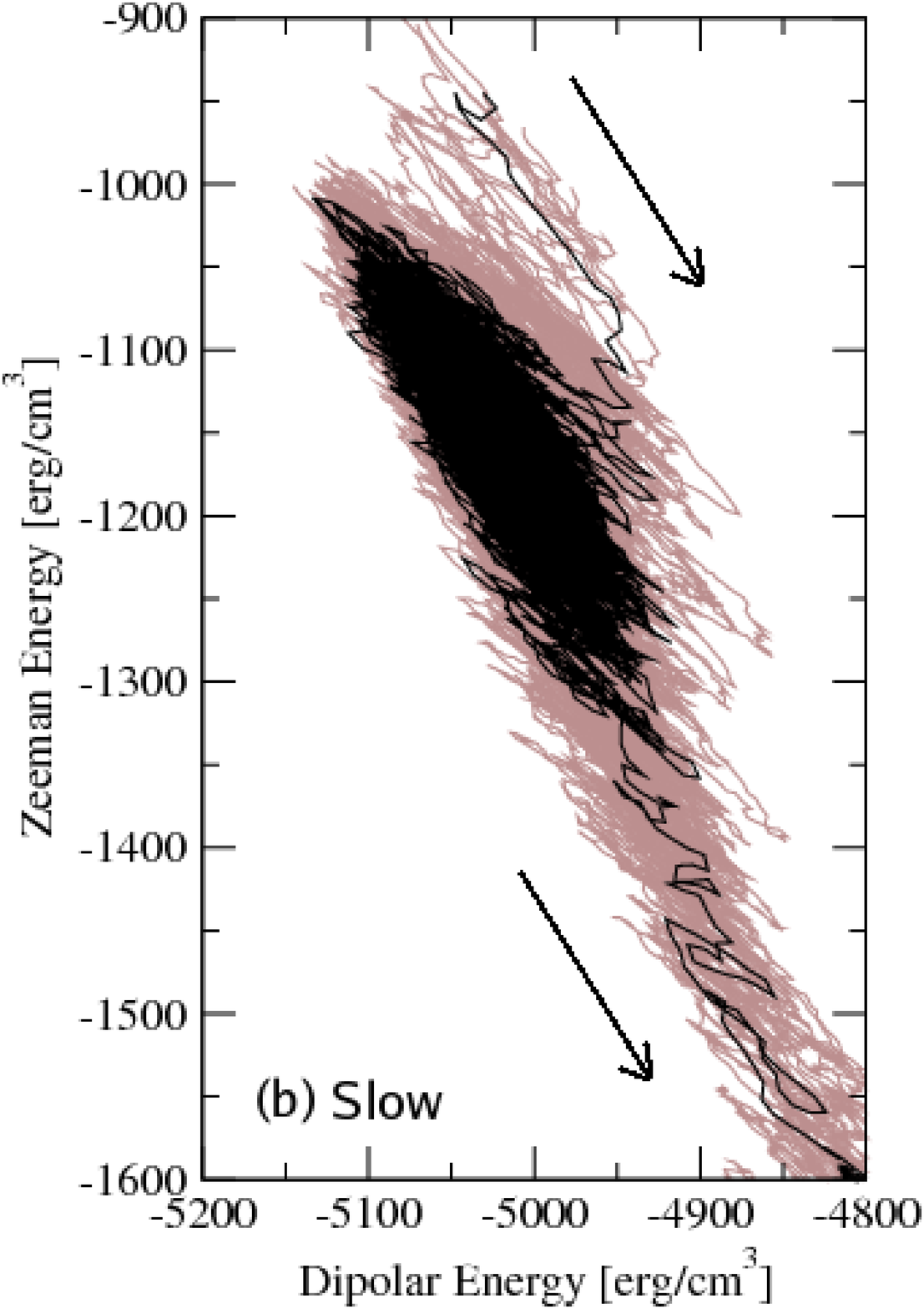} &
\includegraphics[scale=0.17]{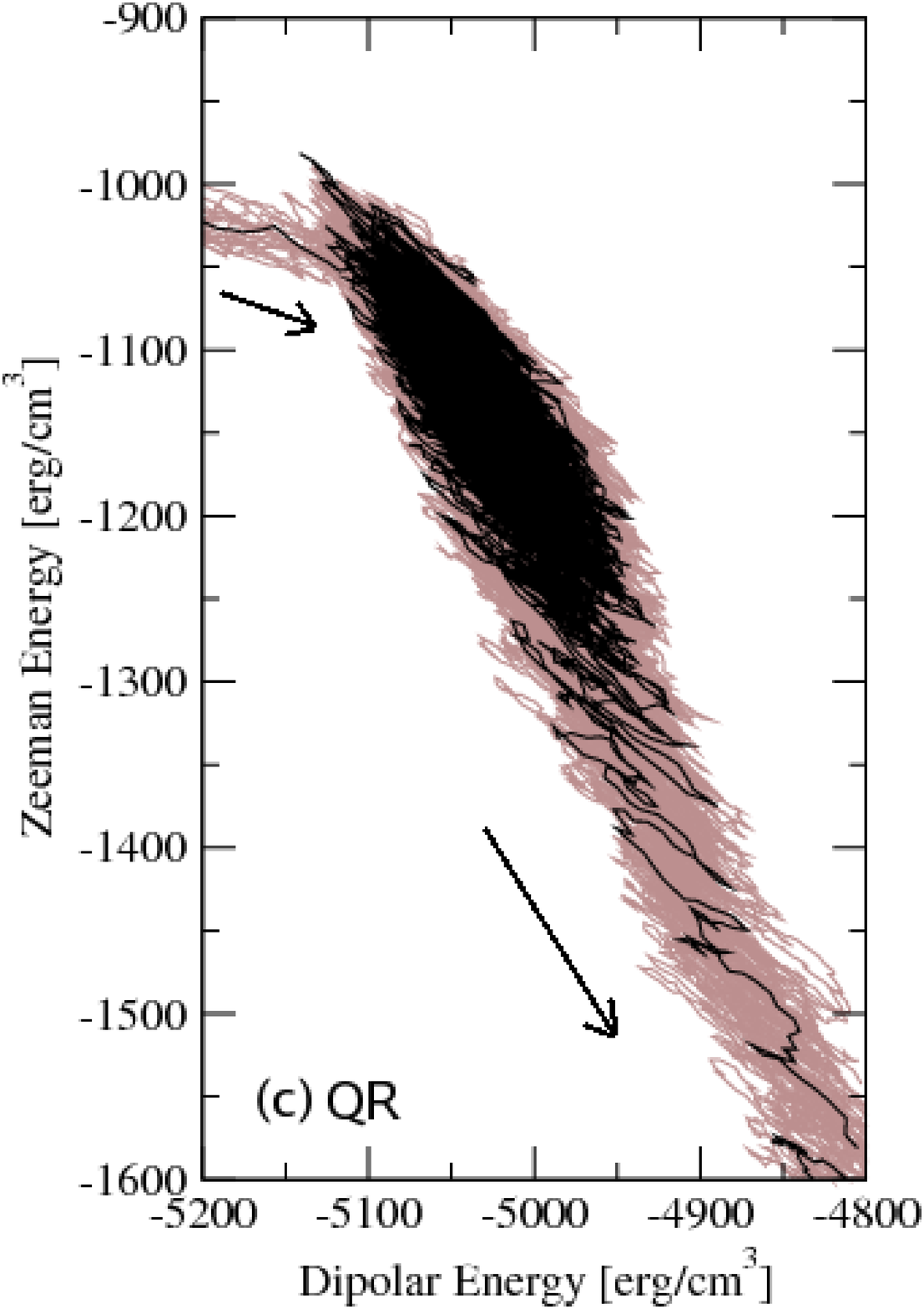}
\end{array}$
\end{center}
\caption
{
(Color online.) Phase plots in the space of dipolar and Zeeman energies for 29 runs belonging to the fast mode~(a), thirteen runs belonging to the slow mode~(b), and 20 runs belonging to the quenched-relaxed trials~(c). 
The lighter background is the collection of all runs belonging to a particular mode, while the darker path represents a single run.
Arrows indicate the average direction of motion of the phase portrait.
}
\label{fig:PP_1}
\end{figure}

Phase plots of the energy also provide information about the behavior of the simulated nanopillar system.
Such plots are shown in Fig.~\ref{fig:PP_1}, with the energies due to dipolar and Zeeman contributions on separate axes.
The collection of all runs belonging to a particular mode are shown in the background of each plot, with a single run overlaid on top.
For the fast~(a) and slow~(b) modes, the simulations begin near the top of the plots, where the density of points is low.
During the initial relaxation of these simulations, both modes evolve down and to the right (i.e., decrease in Zeeman energy and increase in dipolar
energy).
The obvious difference between plots (a) and (b) is the path that each takes near the metastable well ($E^{\mathrm{Z}}\approx-1130$ erg/cm$^3$) and saddle point ($E^{\mathrm{Z}}\approx-1350$ erg/cm$^3$), determined projective dynamics in Section~\ref{PD}.
The slow-mode trajectories proceed to the metastable well, which can be seen in plot~(b) as the large dark region stretching from the center to the top left of the plot, but is absent in plot~(a). 
For the slow mode, the simulation spends most of its time here.
However, the fast-mode events ignore this attractor almost completely, mostly slowing down only near the saddle point.
This is not unreasonable, since the free energy near the saddle point necessarily has a small gradient, and the driving force is therefore weak.
Both modes continue the switching process toward the global free-energy minimum located below the displayed portion of the phase portraits.
These results suggest that the difference between the fast and slow modes is the visitation of the metastable well by the slow mode.
This is further supported by the QR simulations discussed below.

The QR system begins at a point located more negative than $E^{\mathrm{D}}\approx-5200$ erg/cm$^3$ along the dipolar axis of Fig.~\ref{fig:PP_1}(c) (not visible), which represents the common initial configuration for all QR trials.
This initial configuration was found by quenching several simulations belonging to both the slow and fast modes at various values of $E_\mathrm{Z}$ during the evolution of the system.
Quenches initiated at values higher than the saddle point in $E_\mathrm{Z}$ equilibriated to a common $T = 0$ K metastable configuration.
This configuration quickly proceeds to the $T = 20$ K metastable well when thermalized.
Other quenches, which were initiated at values of $E_\mathrm{Z}$ below the saddle point, settled into configurations near the $T = 20$ K final absorbing state.

\begin{figure}[tb]
\centering\includegraphics[scale=0.40]{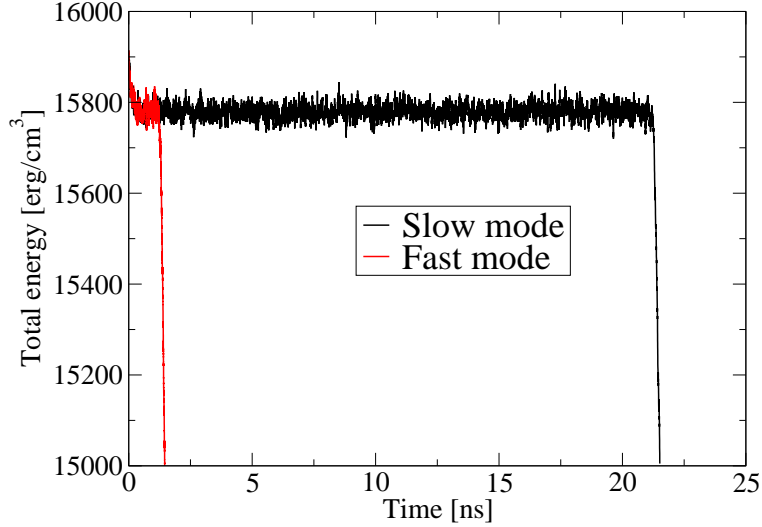}
\caption[]
{
(Color online.) Total energy $E$ as a function of time for a single run belonging to the slow mode and a single run belonging to the fast mode.
Near $t = 0$, the total energy quickly decreases as the system initially relaxes.
After this, the average value of the total energy remains constant, except for small fluctuations, until the saddle point is crossed.
}
\label{fig:TEnergy}
\end{figure}

Since this system necessarily starts in the metastable well, it does not have the same behavior as the fast and slow modes during
the re-thermalization.
However, ignoring this initial relaxation, it closely resembles the phase plot for the slow mode.
The large change with $T$ in the phase-space location of the metastable state of the QR systems is interesting in itself.
It indicates that the entropy $S$, which enters the free energy $F$ as $F = E - TS$, has a large influence on the free energy of the system.
This observation is further supported by the fact that the system remains at a constant total energy $E$ (except for small fluctuations) until the saddle point is crossed, as seen in Fig.~\ref{fig:TEnergy}.

\subsection{Transition Matrix}
\label{TM}

Although the phase space in these simulations is very large ($2N$ dimensions), it appears from the phase plots in Fig.~\ref{fig:PP_1} that the evolution of the system along the Zeeman-energy coordinate, $E^\mathrm{Z}$, can approximately describe the process of magnetization switching.
To investigate the possibility of a one-dimensional description of the switching process, this coordinate is discretized into $200$ equal-sized bins labeled by $i = 1, ..., k$.
We found the results of the following sections to be approximately independent of the axis discretization, and we therefore use this value since it provides numerically stable results for all of the Markov-chain techniques.
Each measurement of $E^\mathrm{Z}$ during the simulation corresponds to a particular bin $i$, and this discretized state can be represented by a unit vector ${\langle}\hat{i}|$ consisting of $1$ in the $i$th position with all other elements equal to $0$.

The matrix, $\mathbf{M}$, of transition probabilities between states is also constructed by sampling the series of $E^\mathrm{Z}$ values during the simulation.
Individual elements of the transition matrix, $M_{ij}$, are obtained by enumerating the single transitions from bin $j$ to bin $i$ corresponding to a time step, $\Delta{t}$, equal to the measurement interval during the simulations, and normalizing such that $\sum_{j}M_{ij} = 1$.
Thus, the probability of going from  ${\langle}\hat{i}|$ to ${\langle}\hat{j}|$ in one time step is given by the matrix element $M_{ij}$ with the result that,
\begin{equation}
{\langle}u(t+\Delta{t})| = {\langle}u(t)|\mathbf{M},
\label{mateq1}
\end{equation}
where ${\langle}u(t)| = \sum_{i}{\langle}u(t)|\hat{i}{\rangle}{\langle}\hat{i}| = \sum_{i}p_{i}(t){\langle}\hat{i}|$ is the row vector representing the probability that $E^{\mathrm{Z}}$ is in bin $i$ at time $t$.
Therefore, $\mathbf{M}$ provides the average change in the state of the system after one time step.~\cite{Novotny:atoadmcmfswdss}
These transition probabilities are found by combining statistics from all individual runs belonging to a particular mode (fast, slow, or QR).

Since the average time evolution of the simulation decreases $E^\mathrm{Z}$, the system begins in state ${\langle}\hat{k}|$ and ends when the absorbing state ${\langle}\hat{1}|$ is reached and $m_z \leq 0$ (the $z$-component of the total magnetization is $\leq 0$).
From this condition, the matrix element $M_{11} = 1$ signifies this absorbing state.
The transition matrix representing this absorbing Markov chain thus has the form,~\cite{Iosifescu:fmpata}
\begin{equation}
\mathbf{M}_{(r+s)\times(r+s)} = \begin{pmatrix}
	\mathbf{I}_{r{\times}r} & \mathbf{0}_{r{\times}s}\\
	\mathbf{R}_{s{\times}r} & \mathbf{T}_{s{\times}s}
\end{pmatrix}.
\end{equation}
Here, $\mathbf{I}$ is an identity matrix which, in general, represents $r$ absorbing states (here, $r=1$).
$\mathbf{R}$ is the recurrent matrix, which describes the probability of moving into the absorbing state from any other state, $\mathbf{0}$ is a null matrix, and $\mathbf{T}$ is the transient matrix, which describes the evolution of the system before absorption.

By this construction, $\mathbf{M}$ is a non-symmetric, regular, non-negative square matrix with different left and right eigenspaces.
The Perron-Frobenius theorem~\cite{Iosifescu:fmpata} provides a general property for this type of matrix, namely that there exists a unique eigenvalue of $\mathbf{M}$ equal to unity, which is larger than the magnitudes of all other eigenvalues of $\mathbf{M}$. 
For our matrix, the left eigenvector associated with the largest eigenvalue of $\mathbf{M}$ is $\langle\lambda_{1}| = \langle\hat{1}| = (1, 0, 0, ...0)$.
This represents the probability distribution characterizing the absorbing state.

\begin{figure}[tb]
\centering\includegraphics[scale=0.40]{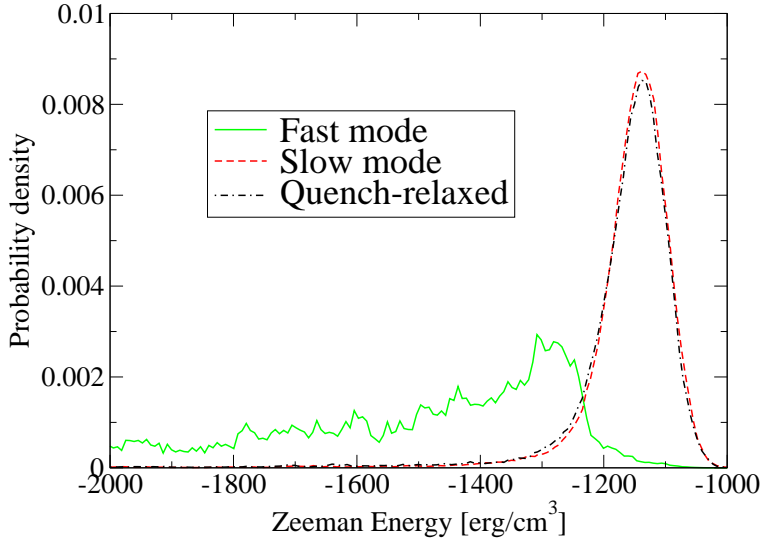}
\caption[]
{
(Color online.) Probability distributions obtained from the eigenvectors corresponding to the second largest eigenvalue for the fast and slow modes and the QR simulations.
In this figure, the Zeeman axis has been discretized into 200 bins to be consistent with the projective dynamics analysis of Sec.~\ref{PD}.
We found the probability densities to be approximately independent of the discretization.
These distributions are constructed from all the data belonging to each mode, as discussed in Section~\ref{TM}.
The slow and QR distributions have a well-defined peak at the location of the metastable free-energy well.
The fast mode, however, is quite wide with its greatest probability near the saddle point.
}
\label{fig:LEVs}
\end{figure}
The other eigenvalues of $\mathbf{M}$ correspond to decaying deviations from the equilibrium distribution since ${\langle}\lambda_{1}|\mathbf{M} = {\langle}\lambda_{1}|$ and 
\begin{equation}
{\langle}\lambda_{\alpha}|\mathbf{M} = \lambda_{\alpha}{\langle}\lambda_{\alpha}|
\label{mateq2}
\end{equation}
with $0 < |\lambda_{\alpha}| \leq 1$ implying that the average probability of being in each non-absorbing bin decreases with each simulation step.
Here ${\langle}\lambda_{1}| = {\langle}\hat{1}|$ denotes the dominant left eigenvector of $\mathbf{M}$, and ${\langle}\lambda_{\alpha}|$ is any other eigenvector of $\mathbf{M}$ with eigenvalue $\lambda_{\alpha}$.
As a consequence of the above, all left eigenvectors of $\mathbf{M}$ except ${\langle}\lambda_{1}|$ will decay to zero under repeated applications of $\mathbf{M}$ since ${\langle}\lambda_{\alpha}|\mathbf{M}^{n} = {\lambda_{\alpha}}^{n}{\langle}\lambda_{\alpha}|$.
Equations~(\ref{mateq1}) and~(\ref{mateq2}) together are used to find the lifetime $\tau$ of the $\alpha$th eigenstate, with the result that,
\begin{equation}
{\langle}\lambda_{\alpha}(t)| = \lambda_{\alpha}{^t}{\langle}\lambda_{\alpha}(0)| \approx e^{-(1-\lambda_{\alpha})t}{\langle}\lambda_{\alpha}(0)|.
\end{equation}
Since ${\langle}\lambda_{2}|$ corresponds to the second longest-lived state of the system, $\tau_{\mathrm{EV}} = 1 / ( 1 - \lambda_2 )$ represents the average lifetime of the metastable state of the system in units of the measurement time resolution.

A probability distribution representing the metastable state can be built from a properly normalized linear combination of ${\langle}\lambda_1|$ and ${\langle}\lambda_2|$, ${\langle}\mathrm{meta}| = {\langle}\lambda_{1}|$ + $b{\langle}\lambda_{2}|$.
Here, the scaling constant $b$ constrains the resulting vector to have zero weight in the absorbing state and ensures that the probability is normalized ($\sum_{i}{\langle}\mathrm{meta}|\hat{i}{\rangle} = 1$).
As can be seen in Fig.~\ref{fig:LEVs}, the slow and QR metastable state probability distributions have a similar shape and exhibit peaks at the location of the metastable free-energy well.
The fast mode, however, does not have a dominant peak in the probability distribution.
Rather, the probability density is close to zero near the metastable free-energy well and spread out along the remainder of the axis, exhibiting a marginally larger density near the location of the free-energy saddle point.
These results reinforce the conclusion that the fast mode switches simply do not fall into the metastable well.

A summary of the lifetimes obtained by the methods discussed above is provided in Table~\ref{table1}.
We expect some bias in the lifetime data due to the artificial cut made when sorting fast and slow modes.
This probably accounts for the large $t_{0}$ for the slow mode.
Trials with lifetimes $<2$ ns are considered fast modes, although there is some probability that they may belong to the slow mode, albeit with a short individual lifetime.
This may be reflected in the differences of the lifetimes of the slow mode and the QR trials.
Differences between ${\tau}_{\mathrm{EV}}$ and the other measurements of the fast-mode lifetimes may be due to statistical error which is not included in the transition matrix associated with ${\tau}_{\mathrm{EV}}$, or that for the fast mode analyzing only a single eigenmode may not be sufficient.
We expect the slow and QR lifetimes obtained from the eigenvalues of the transition matrix to be more robust against this error due to the larger amount of statistics gathered from longer runs.

\begin{table}
\caption
{
Lifetimes in nanoseconds were obtained by the following methods:
$\langle{t}\rangle$ is the empirical mean obtained from the switching times of the actual simulations, and $\sigma_t$ is the corresponding standard deviation.
For the slow and QR modes, characterized by the function $f(t) = \eta(t-t_0)(1/\tau)\exp(-(t-t_{0})/\tau)$, $t_0$ is found from $\min\{$shortest switching time$, \langle{t}\rangle - \sigma_{t}\}$, and $\tau = \langle{t}\rangle - t_0$.
Since the fast mode is not well described by this function, $t_0$ and $\tau$ are not defined for this mode.
${\tau}_{\mathrm{EV}}$ is obtained from the eigenvalues of the transition matrix.
${\tau}_{\mathrm{RT}}$ is found from the residence times of the PD analysis.
}
\begin{tabular*}{0.75\textwidth}%
   {@{\extracolsep{\fill}}cccr}
      & Fast Mode (ns)    & Slow Mode (ns)  &  Quenched Relaxed (ns)\\
\hline  
$\langle{t}\rangle$  & 	1.2	&	20.9	&	14.2	\\
$\sigma_t$		&	0.4	&	13.1	&	10.8	\\
$t_{0}$			&	N/A	&	7.8	&	1.6	\\
$\tau$			&	N/A	&	13.1	&	12.6	\\
${\tau}_{\mathrm{EV}}$  &	0.5	&	21.6	&	13.5	\\
${\tau}_{\mathrm{RT}}$  &	1.2	&	22.7	&	15.2	\\
\end{tabular*}
\label{table1}
\end{table}

\subsection{Projective Dynamics}
\label{PD}

The transition matrix of the previous section accounts for all transitions that may occur from a given bin, potentially with all elements of the submatrices $\mathbf{R}$ and $\mathbf{T}$ non-zero.
However, since the bin sizes are chosen sufficiently large, the transition matrix is tridiagonal.
Under this condition, the one-dimensional description of the magnetization switching becomes a one-step Markov process.~\cite{Novotny:atoadmcmfswdss}
Using this single, coarse-grained variable, the projective dynamics~(PD) method~\cite{Novotny:atoadmcmfswdss, Kolesik:apmfsadolss, Kolesik:pdfmdiim, Brown:pdamr, Kolesik:eldmcs} can be used to measure the growth probabilities $P_\mathrm{G}$ and shrinkage probabilities $P_\mathrm{S}$ of the stable phase along this coordinate.
Once obtained, several properties of the projected free energy $F(E^\mathrm{Z})$ can be measured via $P_\mathrm{G}$ and $P_\mathrm{S}$.

The PD method is implemented as follows.
First, the $E^\mathrm{Z}$ axis is broken into a number of bins, as described for the transition-matrix approach in Sec.~\ref{TM}.
The bin size is determined such that each time step is only capable of moving the system between adjacent bins.
Histograms of $P_\mathrm{G}(E^\mathrm{Z})$, $P_\mathrm{S}(E^\mathrm{Z})$, and the probability to stay in the same bin $P_\mathrm{N}(E^\mathrm{Z})$, keep record of changes along the axis and are updated at each measurement.
Once completed, the histogram is then normalized so that $P_\mathrm{G}(E^\mathrm{Z}) + P_\mathrm{S}(E^\mathrm{Z}) + P_\mathrm{N}(E^\mathrm{Z}) = 1$.

For the present nanopillar simulations, we attempted to use as the slow (binned) variables the Cartesian components of the total magnetization, as well as other contributions to the total energy.
However, the Zeeman energy provided the best results for the projective-dynamics technique.
This is reasonable since it provides the closest correspondence to the path observed in Fig.~\ref{fig:PP_1}.

\begin{figure}[tb]
\centering\includegraphics[scale=0.40]{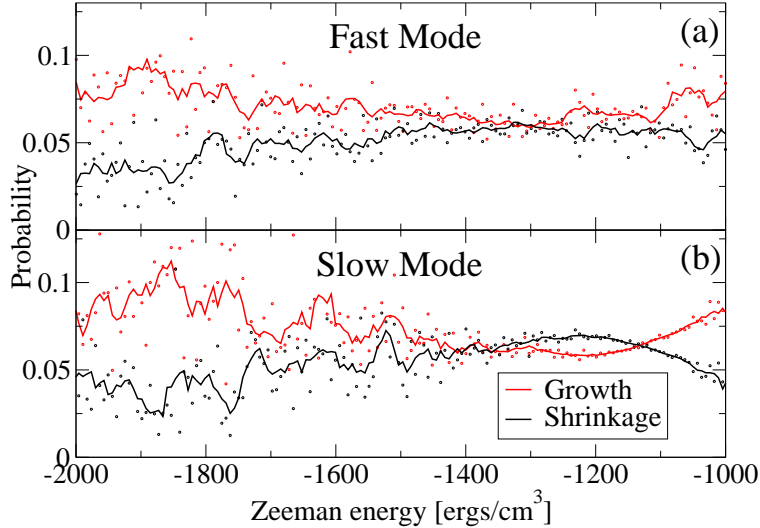}
\caption[]
{
(Color online.) Projective Dynamics results for the two modes, fast (a) and slow (b).
The slow mode exhibits clear crossings of the growth and shrinkage probabilities, indicating extrema in the free energy, corresponding to the metastable well (right) and the saddle point (left).
Conversely, the probabilities for the fast mode only appear to overlap in this same region.
The solid lines are five-point running averages.
}
\label{fig:PD_1}
\end{figure}

Points along $E^\mathrm{Z}$ where $P_\mathrm{G} = P_\mathrm{S}$ define local extrema of the free energy.
This is true since $P_\mathrm{G} > P_\mathrm{S}$ implies ${dF}/{dE^\mathrm{Z}}>0$ and $P_\mathrm{S} > P_\mathrm{G}$ implies
${dF}/{dE^\mathrm{Z}}<0$.
For our system, which has a single metastable free-energy well, these extrema represent the location of the metastable well, the saddle point,
and the true equilibrium in the free energy.
The latter is not observed in our simulations due to the cut-off at $m_z \leq 0$.

Figures~\ref{fig:PD_1} (a) and (b) show PD plots for the fast and slow modes, respectively.
Each plot contains the data of all the runs belonging to each mode (points), which have then been smoothed using a five-point running average
(solid curves).
The location of the metastable free-energy well for the slow-mode switches (first crossing from the right in Fig.~\ref{fig:PD_1} (b), $E^\mathrm{Z} \approx -1130$ ${\mathrm{erg}}/{\mathrm{cm^3}}$) coincides with the peaks present in the metastable probability distribution, Fig.~\ref{fig:LEVs}, of Sec.~\ref{TM}. 
Further left, the second crossing in Fig.~\ref{fig:PD_1} (b) indicates the saddle point in the free energy, which is located at $E^\mathrm{Z} \approx -1350$ ${\mathrm{erg}}/{\mathrm{cm^3}}$
The locations of both the metastable well (first crossing from the right) and the saddle point (second crossing from the right) are obvious for 
the slow mode.
This is expected since the results of Sec.~\ref{TM} indicate true metastable behavior.

\begin{figure}[tb]
\centering\includegraphics[scale=0.40]{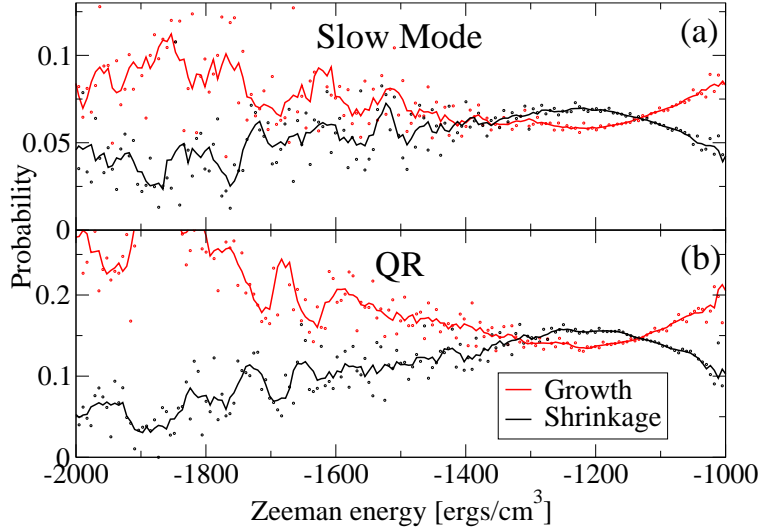}
\caption[]
{
(Color online.) Projective Dynamics results for the two modes, slow and the quenched-relaxed (QR).
It is easy to see that these simulations share the same statistical behavior in this region, including nearly identical crossing values (which are locations of the extrema of the free energy).
As with Fig.~\ref{fig:PD_1}, the solid lines represent five-point running averages.
}
\label{fig:PD_2}
\end{figure}

From our data, the fast mode has an overlap of $P_\mathrm{G}$ and $P_\mathrm{S}$ in the range from $E^\mathrm{Z} = -1350$ ${\mathrm{erg}}/{\mathrm{cm^3}}$ to $E^\mathrm{Z} = -1250$ ${\mathrm{erg}}/{\mathrm{cm^3}}$.
It is not possible to determine if crossings occur in this region.
This could be because crossings exist but are obscured by the small amount of statistics available, since less time is spent in this region than for the slow mode.
It could also be that no true crossings occur.
In this case fast swithches proceed through a region of the free-energy landscape with a small, but negative slope.
Along with the results of the metastable probability distributions, this indicates the absence of a metastable state in the fast mode.
These results agree with the probability distributions (Fig.~\ref{fig:LEVs}) obtained from the transition matrices of the previous section.

The PD results of the quenched-relaxed system can be compared to the slow modes.
As seen in Fig.~\ref{fig:PD_2}, the QR system and the slow-mode probabilities near the metastable minimum cross at nearly identical values of $E_\mathrm{Z}$.
Both PD and the metastable probability distributions imply that, once in the metastable well (the slow and QR decays), the system must be thermally activated to overcome the free-energy saddle point.
However, if the metastable well is avoided (the fast mode), the system only has to traverse a relatively flat free-energy landscape to switch.

Lifetimes can be calculated easily from the results of the PD technique.~\cite{Novotny:atoadmcmfswdss}
The residence time for each bin $h(i)$ is defined as the average time spent in state $i$. 
In addition, $P_\mathrm{G}(i)h(i)$ is the average number of times the system moves from state $i$ to $i-1$ (the absorbing state is state $1$).
If the system is to reach the absorbing state exactly once then
\begin{equation}
P_\mathrm{G}(i)h(i) - P_\mathrm{S}(i-1)h(i-1) = 1
\end{equation}
must be true.
Since the absorbing state has no shrinking probability, $P_\mathrm{S}(1) = 0$, the residence time of the previous state can be found by $h(2) = 1/P_\mathrm{G}(2)$.
This leaves an iterative solution for the remaining states,
\begin{equation}
h(i)=\frac{1+P_\mathrm{S}(i-1)h(i-1)}{P_\mathrm{G}(i)}.
\end{equation}
Consequently, the average lifetime of the process is the sum of the residence times, $\tau_{\mathrm{RT}} = \sum_i h(i)$.
The lifetimes obtained from this method are included in Table~\ref{table1}.

\section{Conclusions}
\label{S}

We have studied the magnetization-switching properties of a simulated iron nanopillar motivated by experimental research.~\cite{Kent:science262, Wirth:mbonsip}
Under the realistic physical conditions described here, we found the presence of more than one characteristic switching time, which were labeled as ``fast'' and ``slow'' modes.
Through phase portraits and numerical results provided by transition matrices and the projective dynamics method, differences between these
two modes were identified.
Our results indicate that the fast mode is associated with switching dynamics that do not carry the system through a metastable well.
This idea is supported by data from separate simulations in which the system was quenched to 0 K while near the metastable well and then rethermalized.
The quenched-relaxed simulations also indicate that the entropy provides a large contribution to the free energy of the system.

Using transition matrices obtained from the average behavior of each mode, we also constructed projective dynamics plots and metastable probability distributions which further provide evidence that the fast mode does not have a metastable free-energy well.
In addition, lifetimes were obtained by measurement, fitting to the cumulative distribution, sub-dominant eigenvalues, and residence times from projective dynamics and reported in Table~\ref{table1}.
These results support further evidence of the existence of multiple switching paths which may be verified experimentally, and may have important ramifications for technological applications that rely on a single, consistent switching mode such as memory devices.
We hope to extend the scope of these results in future work by measuring the dependence of the fast- to slow-mode ratio as a function of the applied field angle and temperature.

\section{Acknowledgments}

We gratefully acknowledge useful conversations with S. von Moln\'{a}r, Steffen Wirth, and J. M\"{u}ller.
This work was supported in part by NSF grant No. DMR-0444051 and by Florida State University through the Center for Materials Research and Technology, the School of Computational Science, and the National High Magnetic Field Laboratory.
G. Brown was supported by the Division of Scientific User Facilities, U.S. Department of Energy.

\bibliography{umag}

\end{document}